\documentclass[runningheads]{llncs}
\usepackage[T1]{fontenc}
\usepackage{graphicx}
\usepackage{cite}
\usepackage{amsmath}
\usepackage{url}
\usepackage{enumitem}
\usepackage{subfigure}
\usepackage{float}
\usepackage[misc]{ifsym} 
\begin{document}
\title{Small Tunes Transformer: Exploring Macro \& Micro-Level Hierarchies for Skeleton-Conditioned Melody Generation}
\titlerunning{Small Tunes Transformer}
%
\author{Yishan Lv \and
Jing Luo \and
Boyuan Ju \and Xinyu Yang$^{(\textrm{\Letter})}$}
\authorrunning{ Y. Lv et al.}
%
\institute{School of Computer Science and Technology, Xi'an Jiaotong University, Xi'an, China\\
\email{\{yisan,luojingl,juboyuan\}@stu.xjtu.edu.cn, yxyphd@mail.xjtu.edu.cn}}
\maketitle              
\begin{abstract}

Recently, symbolic music generation has become a focus of numerous deep learning research. Structure as an important part of music, contributes to improving the quality of music, and an increasing number of works start to study the hierarchical structure. In this study, we delve into the multi-level structures within music from macro-level and micro-level hierarchies. At the macro-level hierarchy, we conduct phrase segmentation algorithm to explore how phrases influence the overall development of music, and at the micro-level hierarchy, we design skeleton notes extraction strategy to explore how skeleton notes within each phrase guide the melody generation. Furthermore, we propose a novel Phrase-level Cross-Attention mechanism to capture the intrinsic relationship between macro-level hierarchy and micro-level hierarchy. Moreover, in response to the current lack of research on Chinese-style music, we construct our Small Tunes Dataset: a substantial collection of MIDI files comprising 10088 Small Tunes, a category of traditional Chinese Folk Songs. This dataset serves as the focus of our study. We generate Small Tunes songs utilizing the extracted skeleton notes as conditions, and experiment results indicate that our proposed model, Small Tunes Transformer, outperforms other state-of-the-art models. Besides, we design three novel objective evaluation metrics to evaluate music from both rhythm and melody dimensions.

\end{abstract}

\section{Introduciton}
Music stands as a treasure within human civilization. In recent years, music generation has become a focus of deep learning research. Many sequence models has been employed to generate symbolic music\cite{r2,r3,r4,r5}. Following the introduction of Music Transformer\cite{r20}, which utilizes a transformer-based architecture for music generation, several Transformer-based models have made a progress in generating complete melodies\cite{r9,r22,r19}.

Structure is of great significance to music, recently, plenty of works start to study the hierarchical structural features within the music. \cite{r31} studies the phrase-level hierarchy of music, \cite{r9,r32} study the bar-level hierarchy of music, \cite{r26} studies the phrase \& bar-level hierarchies of music. While beneath the bar-level hierarchy, there exists a micro-level hierarchy in which skeleton notes play an important role. In this study, we delve into the hierarchical structure, exploring the intrinsic relationship among macro-level hierarchy and micro-level hierarchy.

\begin{figure*}[tbp]
    \centering
    \includegraphics[width=0.9\textwidth]{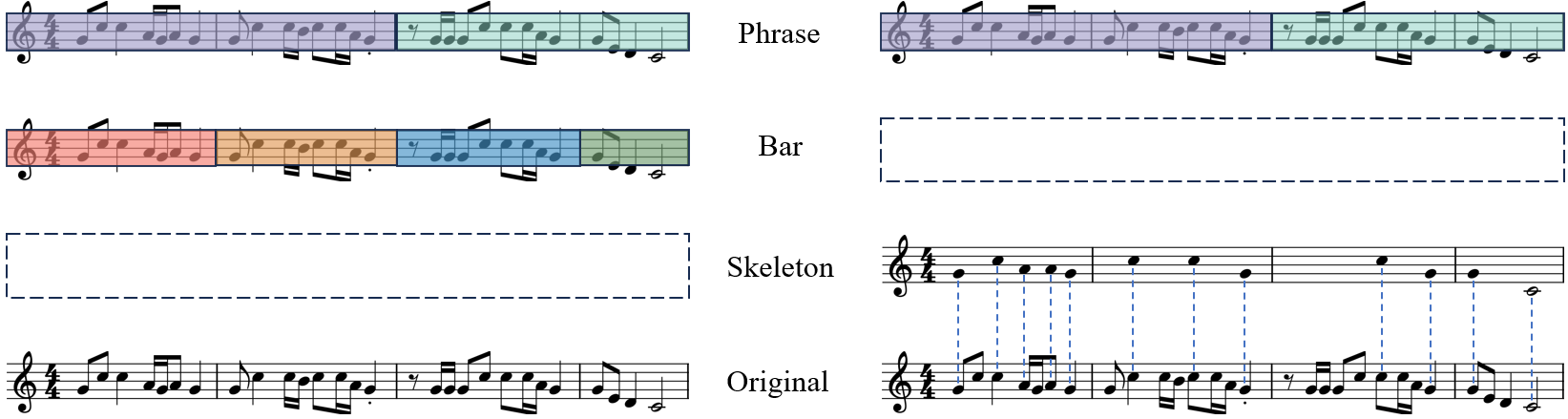}
    \caption{Two multi-level hierarchies: phrase \& bar-level hierarchies (left) and our macro \& micro-level hierarchies (right). The dashed boxes indicate levels that are not considered in the respective hierarchies.}
    \label{fig:multi-level}
\end{figure*}

Figure \ref{fig:multi-level} illustrates the difference between the phrase \& bar-level hierarchies and our macro \& micro-level hierarchies. Specifically, a melody comprises several phrases, with each phrase comprising several bars. In the phrase \& bar-level hierarchies, the bar serves as the fundamental structure unit. Within a bar there are several notes, among which some play a crucial role in guiding the melody generation. These significant notes, known as skeleton notes, are extracted to establish the micro-level hierarchy in our macro \& micro-level hierarchies. For Chinese Folk Songs, most phrases are relatively short and the distinction between phrase-level and bar-level hierarchies is not that obvious, so we examine phrases instead of bars as the macro-level hierarchy.

Accordingly, we conduct phrase segmentation on the melody at the macro-level hierarchy, and design a skeleton notes extraction strategy within each phrase at the micro-level hierarchy. Especially, we define a new type of skeleton note for Chinese Folk Songs. Building upon this, we propose a novel Phrase-level Cross-Attention mechanism, which enables the model a deep understanding of musical features from both macro-level and micro-level hierarchical structures.

We construct our own dataset: Small Tunes\footnotemark[1] Dataset (STD), and utilize it to train our model: Small Tunes Transformer (STT). Utilizing the extracted skeleton notes as conditions, STT is capable of generating Small Tunes songs with clear structure and captivating melody. We design 3 novel metrics to evaluate the quality of music from pitch and rhythm dimensions. The experiment results indicate that STT outperforms other state-of-the-art models on all 5 subjective evaluation metrics and 5 out of 6 objective evaluation metrics. Besides, we add 6 ablative groups to study the impact of changes in macro-level and micro-level hierarchies on music generation, thereby exploring the hierarchical structural features within music.

\footnotetext[1]{\textsl{Small Tunes}, as known as \textsl{XiaoDiao} in Chinese phonetics, is a category of Chinese Folk Songs. For details, see \url{https://en.chinaculture.org/library/2008-01/11/content_71371_3.htm}}

Our main contributes can be summarized as follows:
\begin{itemize}[label=\textbullet]

\item We propose STT, a Transformer-based model, incorporating the novel Phrase-level Cross-Attention mechanism, to explores the hierarchical structures within music from both macro-level and micro-level hierarchies.
\item We design three objective evaluation metrics: Theme Pitch Corresponding (TPC) and Theme Rhythm Corresponding (TRC) evaluate the coherence corresponding to the theme from pitch and rhythm dimensions, and Pentatonic Scale Consistency (PSC) evaluates consistency in a Chinese-style scale dimension.
\item We construct our own dataset: STD, a large-scale dataset containing 10088 MIDI files, covering almost all recorded Small Tunes songs in China.

\end{itemize}

\section{Related Work}

Music Transformer \cite{r20} is the first work to utilize the Transformer-based architecture to generate music with coherent structure. Drawlody\cite{r37}, a music generation system, composes music by converting a user-input melody curve into melody. MusicVAE \cite{r4} utilizes a hierarchical decoder to generate music with long-term structure. WuYun \cite{r11} leverages music theory to prioritize the generation of structurally important notes as the skeleton, gradually filling in ornamental notes to complete the melody. While WuYun effectively generates coherent melodies, it lacks consideration for structural features within music. In this paper, we build upon the principles of WuYun to explore the intrinsic relationship between macro-level and micro-level hierarchies in music.

In recent years, an increasing number of works have focused on the structural features of music. These studies can be categorized based on their exploration of intrinsic structural hierarchies into four types: phrase-level, bar-level, phrase \& bar-level, and others.
1) phrase-level: MusicFrameworks \cite{r8}, a Transformer-LSTM architecture, processes music sequences by incorporating chord, melody, and rhythm features. \cite{r39} generates music by imitating the structure, melody, and style of a given seed song. \cite{r31} explores the form, harmony, and texture features to enhance the structure within music. Theme Transformer \cite{r34} centers on theme-based conditioning, generating music using thematic material as the condition.
2) bar-level: Melons \cite{r9}, a Transformer-based music generation model, represents music sequences as graphs based on eight custom-defined structural types. Popmnet \cite{r32} generates pop music with a well-organized structure by establishing relationships of repetition and sequence between all bars.
3) phrase \& bar-level: Hyperbolic Music Transformer \cite{r38} enhances the structure of music by leveraging hyperbolic theory. \cite{r28} utilizes a data-driven approach to analyze the structure of symbolic music. \cite{r26} proposes the Phrase and Bar Countdown events to study the phrase \& bar-level hierarchies within music.
4) Others: \cite{r24b} explores repetitive patterns at the motif-level. \cite{r10} progressively expands a music fragment into a complete melody across the motif, phrase, and section levels. \cite{r33} explores structural elements at the note, chord, and section levels in music to enhance its quality.

Most of the aforementioned works concentrate on music generation within Western music genres such as Western pop music, while research on Chinese-style music, especially Chinese Folk Songs, remains relatively limited. Although some researchers have employed sequence-to-sequence models for Chinese-style music generation, such as MG-VAE \cite{r7} for regional-style Chinese Folk Songs composition, \cite{r12} generates melody and arrangement for Chinese pop-style songs.

\section{Method}
\subsection{Phrase Segmentation}

The structure of Chinese Small Tunes is unique, often presenting orderly structural patterns. The distinctive hierarchical structure in Chinese Small Tunes reflects traditional style of Chinese Folk Songs. Most of the phrases within Small Tunes are relatively short, and thus we examine the phrases as macro-level hierarchy.

We dedicate to produce the accurate phrase segmentation of Small Tunes, which is significant to explore the intrinsic structural features within music. 
We apply a deep learning method to get phrase segmentation. The model architecture we select is a convolutional neural network with conditional random field \cite{r29}, and 1168 labeled Chinese Folk Songs in public data set Essen Folksong Database are used to train the model. Then the phrase segmentation of each song in our dataset can be produced using the trained model. The phrase segmentation of a song is defined as $S=\{s_1,s_2,\ldots,s_n\}$, where $n$ is the length of sequence, and for instance, $s_i=k$ indicates that the $i^{th}$ note belongs to the $k^{th}$ segment.

\subsection{Skeleton Extraction}

A melody consists of structural notes and ornamental notes, these structural notes, called skeleton \cite{r30}, is the underlying framework of a full melody. Based on the melodic skeleton, a full-fledged melody can be composed by filling into ornamental notes. The skeleton notes, which tend to be more prominent in auditory perception, are selected as the micro-level hierarchy for our study.

Skeleton notes can be divided into pitch and rhythm dimensions. One skeleton note extracted from the pitch dimension contributes to the stability and harmony, while one from the rhythm dimension is of importance of the rhythm of melody development.

\begin{figure}[tbp]
    \centering  
    \subfigure[\label{fig:subfig1}]{
        \includegraphics[width=0.3\textwidth]{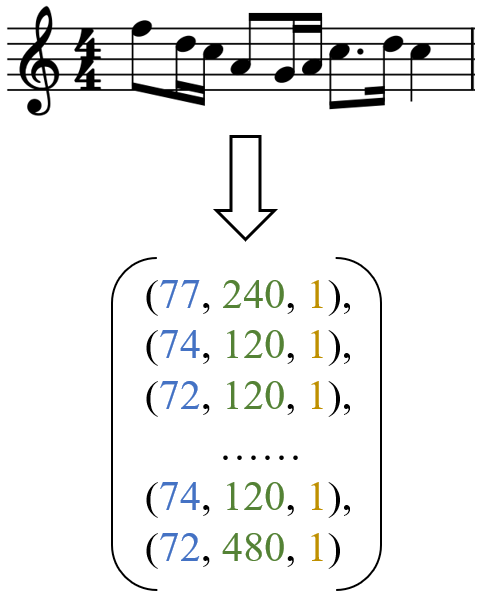}}
    \hspace{0.1\textwidth}
    \subfigure[\label{fig:subfig2}]{
        \raisebox{1.5cm}{\includegraphics[width=0.3\textwidth]{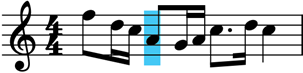}}}
    \caption{(a) An example of music representation: For instance, the first note will be represented as ($77$, $240$, $1$). (b) A piece of \textsl{Molihua}, a famous Chinese Folk Song. The blue-colored $A4$ note, followed by a passing note and returning to $A4$, will be selected as a Small Tunes Trembling Note. }
\end{figure}

For the pitch dimension, we define a Small Tunes Trembling Tote, which often occurs in the Chinese Small Tunes, featuring traditional Chinese style. The Small Tunes Trembling Note starts and ends with the note which has the same pitch, among them there exists some other ornamental notes with shorter duration. Figure~\ref{fig:subfig2} shows one piece of a famous Chinese Folk Song \textsl{Molihua} (or \textsl{Jasmine Flower}) as example.

\begin{figure*}[tbp]
    \centering
    \includegraphics[width=0.93\textwidth]{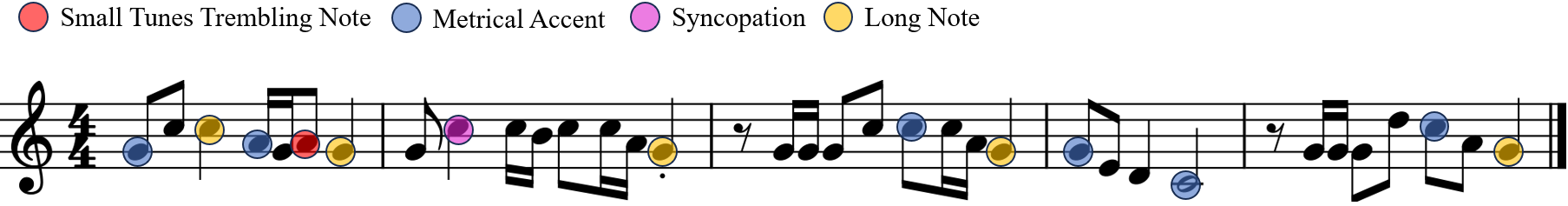}
    \caption{An example of skeleton extraction. The skeleton notes consist of Small Tunes Trembling Note, Metrical Accent, Syncopation and Long Note.}
    \label{fig:skeleton_extraction}
\end{figure*}

For the rhythm dimension, we select three types of skeleton notes according to \cite{r11}, which are metrical accent, syncopation, and long note. After conducting phrase segmentation on a single song, we extract the skeleton notes from each phrase, thereby obtaining the skeleton note sequence. Figure \ref{fig:skeleton_extraction} illustrates an example of skeleton extraction result.

\subsection{Music Representation}
REMI \cite{r15} is a a widely used method for symbolic music representation. However, we utilize a triplet format of $\{pitch, duration, segment\}$ instead of REMI to represent symbolic music sequences for the following reasons: 1) The REMI representation results in an excessively long input sequence, complicating melody modeling. 2) Tokens such as \textit{bar} and \textit{position} in REMI appear irregularly at the beginning or middle of sequences, disrupting the alignment between skeleton notes and full notes sequences during Phrase-level Cross-Attention (as discussed later). Conversely, the triplet format, which includes only \textit{pitch}, \textit{duration}, and \textit{segment} attributes, represents each note as a single token after concatenation. This ensures a one-to-one correspondence between the skeleton notes sequence and the full notes sequence during Phrase-level Cross-Attention, thereby enhancing modeling efficiency.

The \textit{pitch} and \textit{duration} values are obtained directly, while the \textit{segment} value is derived from the outcome of phrase segmentation. After being converted into the digital format, the symbolic music token sequence can be fed into the model as input. Figure~\ref{fig:subfig1} illustrates the music representation.

For pitch sequence $P:\{p_1,p_2,\ldots,p_n\}$, duration sequence $D:\{d_1,d_2,\ldots,d_n\}$, segment sequence $S:\{s_1,s_2,\ldots,s_n\}$. $P, D, S\in R^{n\times 1}$, we embed them as $P_{emb},D_{emb},S_{emb}\in R^{n\times d_{model}}$ where $d_{model}$ represents the embedding dimension. Then we utilize a fusion layer to merge the pitch, duration and segment information, resulting in what we denote as Music Fusion (MF) in Equation \ref{eq:MF}, where $W_{MF}$ represents a trainable linear, and $\oplus$ is a vector concatenation operation.
\begin{equation}
    MF = W_{MF} \cdot (P_{emb} \oplus D_{emb} \oplus S_{emb})
    \label{eq:MF}
\end{equation}

The positional encoding is illustrated in Equation \ref{eq:PE}, $PE_i$ is the original positional encoding of transformer where $I=\{0,1,\ldots,n-1\}$ represents the index of the music sequence, besides, we propose an additional positional encoding $PE_s$ to embed the phrase segment $S:\{s_1,s_2,\ldots,s_n\}$.
\begin{equation}
PE = PE_i + PE_s
\label{eq:PE}
\end{equation}
Now, the input of encoder and decoder block is as follows:
\begin{equation}
input = MF + PE
\end{equation}

\subsection{Model Architecture}

We model a song from macro-level and micro-level hierarchies. At the macro-level hierarchy, a Small Tunes song consists of multiple phrases, which intricately interweave and connect with each other. At the micro-level hierarchy, skeleton notes within each phrase play a pivotal role in guiding the melody generation. In order to better study the intrinsic features among these hierarchical structures, we propose a novel Phrase-level Cross-Attention.

The skeleton notes sequence input of encoder block and the full notes sequence input of decoder block are denoted as $G_{input}$ and $H_{input}$ respectively. After being processed by the encoder block, $G_{input}$ serves as the key and value inputs for the Phrase-level Cross-Attention in decoder block, denoted as $G^{'}:\{g_1,g_2,\ldots,g_m\}\in R^{m\times d_{model}}$, and after being processed by the Masked Relative Self-Attention\cite{r20} and Add \& Norm layer, $H_{input}$ serves as the query input for the Phrase-level Cross-Attention, denoted as $H^{'}:\{h_1,h_2,\ldots,h_n\}\in R^{n\times d_{model}}$. Where $m$ and $n$ are the length of skeleton notes sequence and the length of full notes sequence respectively, and $d_{model}$ is the embedding dimension. The query ($Q$), key ($K$) and value ($V$) are show as Equation \ref{eq:qkv}, where $W_Q$, $W_K$, $W_V$ are three trainable linear layers.
\begin{equation}
    Q, K, V = W_Q \cdot H^{'}, W_K \cdot G^{'}, W_V \cdot G^{'}
    \label{eq:qkv}
\end{equation}

We design a Phrase-level Mask Matrix to ensure that the melody generation of one phrase only attends the skeleton notes within the same phrase, thereby the skeleton notes can guide the melody generation of the corresponding phrase. For explanation purposes, we provide an example as follows. Given the $k^{th} (k\in 1,2,\ldots)$ phrase, after performing phrase segmentation operations as mentioned earlier, we obtain the phrase segmentation labels: $S^g:\{s_1^g,s_2^g,\ldots,s_m^g\}$ for the skeleton notes sequence and $S^h:\{s_1^h,s_2^h,\ldots,s_n^h\}$ for the full notes sequence. Based on this result, we can extract the skeleton notes subsequence $G_k^{'}: \{g_i,g_{i+1},\ldots,g_j\}$ and the full notes subsequence $H_k^{'}: \{h_p,h_{p+1},\ldots,h_q\}$ within the $k^{th}$ phrase, according to Equation \ref{eq:phrase_mask}.
\begin{equation}
    s_i^g = s_{i+1}^g = \dots = s_j^g = k = s_p^h = s_{p+1}^h = \dots = s_q^h
    \label{eq:phrase_mask}
\end{equation}
Where $i$, $p$ are the index of the first note in one phrase and $j$, $q$ are the index of the last note.
Furthermore, after obtaining the index $i$, $j$, $p$ and $q$, the $k^{th}$ block matrix can be represented as Equation \ref{eq:mask_matrix}, where $r$ stands for row, $c$ stands for column, $0$ represents no masking required while $-\infty$ indicates masking.
\begin{equation}
M^{k}=\left\{
\begin{aligned}
& \hfill\phantom{00} 0, && p \leq r \leq q \text { and } i \leq c \leq j \\
& -\infty, && \hfill\phantom{p \leq r \leq q}\hfill\text {others}
\end{aligned}
\right.
\label{eq:mask_matrix}
\end{equation}

Performing the same operation on each phrase yields a total of $n_p$ block matrices, where $n_p$ is the number of phrases. Combining these matrices yields the Phrase-level Mask Matrix $M$.

\begin{figure}[tbp]
    \centering
    \includegraphics[width=0.7\textwidth]{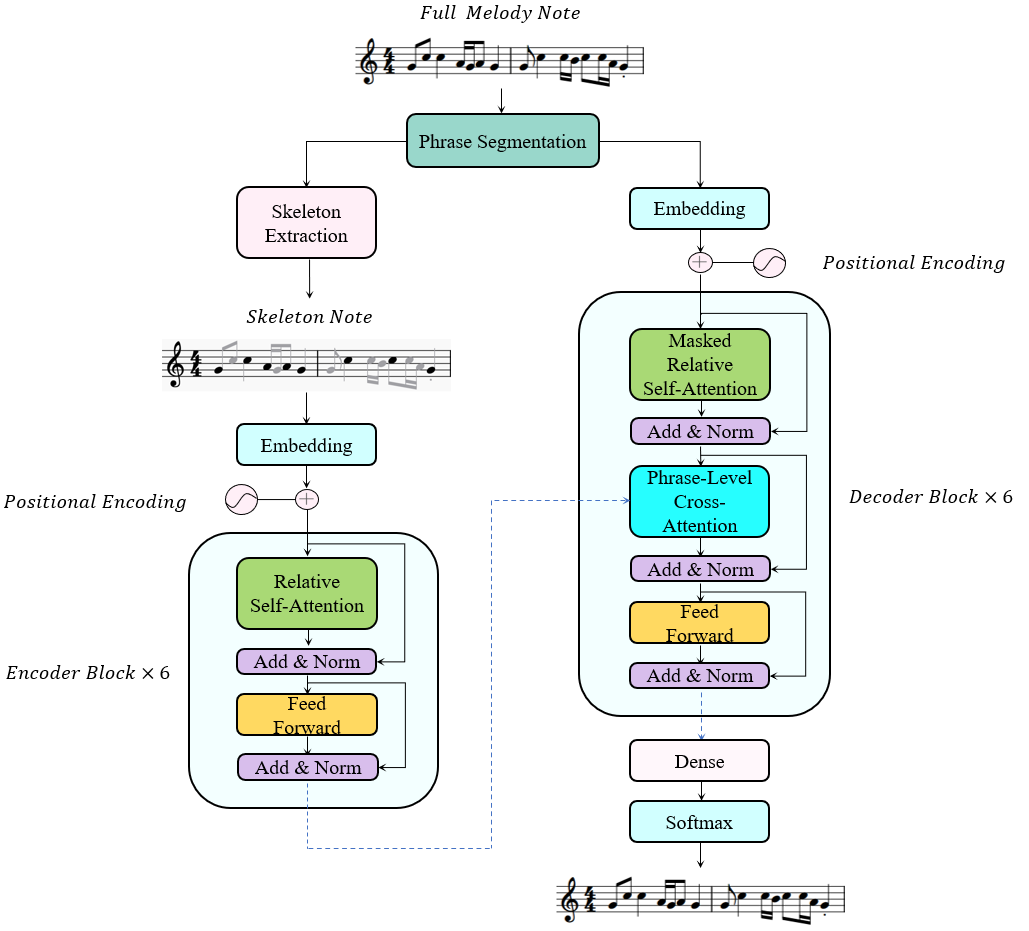}
    \caption{Architecture of Small Tunes Transformer}
    \label{fig:Small Tunes Transformer}
\end{figure}

Finally, the output of Phrase-level Cross-Attention can be obtained as Equation \ref{eq:attention}. Figure \ref{fig:Small Tunes Transformer} illustrates the architecture of our model.
\begin{equation}
    Att(Q,K,V,M) = softmax(\frac{Q\cdot K^T}{\sqrt{d_{model}}} + M) \cdot V
    \label{eq:attention}
\end{equation}

\begin{figure}[tbp]
    \centering
    \includegraphics[width=0.65\textwidth]{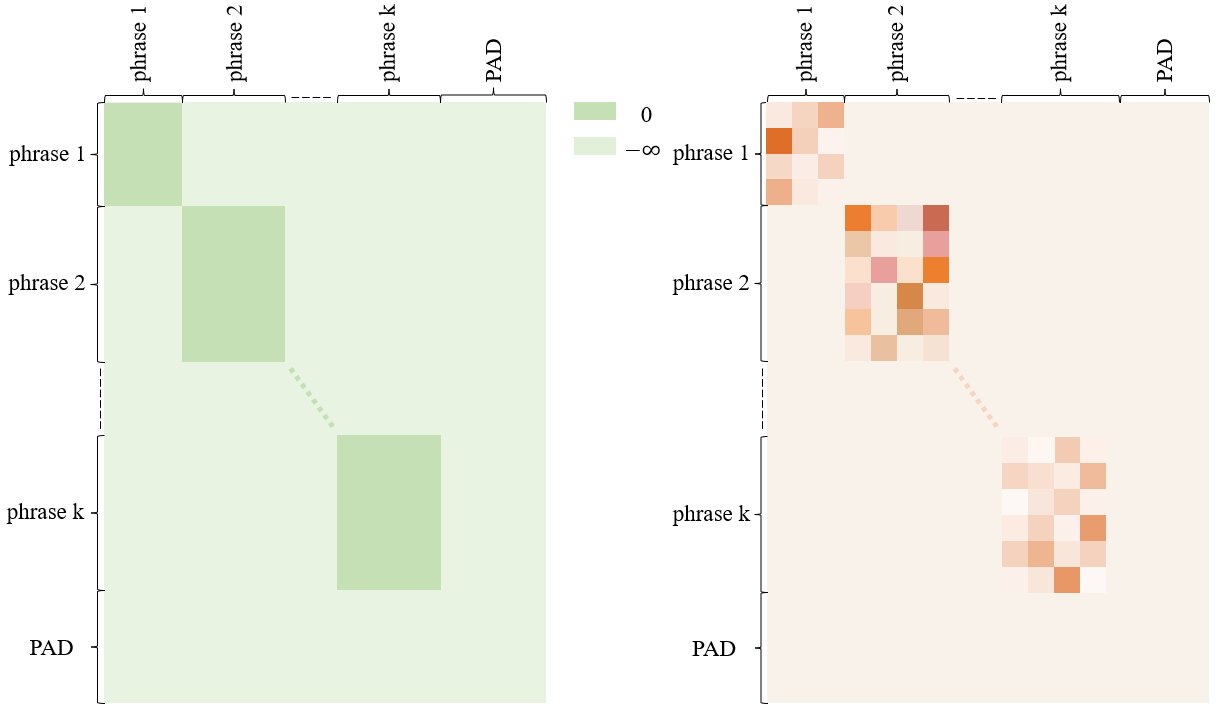}
    \caption{Phrase-level Mask Matrix (left) and attention weights (right)}
    \label{fig:Phrase-Level Mask}
\end{figure}

\section{EXPERIMENT}

\subsection{Experiment Settings}

\subsubsection{Dataset.}
There has been abundant research on Western music genres like classical
and pop music, while studies on Chinese-style songs remain relatively limited.
Chinese Folk Songs, a unique music genre of Chinese-style songs,  with strong
regional characteristics\cite{r35, r36},  captivating melody and richest numbers, can be traced back to the \textsl{Classic of Poetry} (or \textsl{Shijing}) over 3,000 years ago. Small Tunes\footnotemark[1], a category of Chinese Folk Songs, is popular among towns or countries and is characterized by fixed melody and lyrics, orderly structure, and subtle, melodious tunes. Small Tunes serve as the focus of our study.

We construct our dataset, named the Small Tunes Dataset\footnotemark{} (STD), a large-scale collection of 10088 Small Tunes songs. STD encompasses almost all recorded Small Tunes songs from 31 provinces in China, each meticulously transcribed into MIDI format by us. For model training, we select songs with a time signature denominator of 4.

\footnotetext{\url{https://chinglohsiu.github.io/files/MGD.html}}

\subsubsection{Baseline Models.}
In order to explore the advantages of the model architecture, we select three models as our baseline models:
\begin{itemize}[label=\textbullet]
\item \textbf{Music Transformer} (MT), which is the first Transformer-based model to generate symbolic music\cite{r20}.
\item \textbf{WuYun}, which uses the skeleton notes as a condition but lack of any segment information\cite{r11}.
\item \textbf{Music Transformer with Phrase and Bar Countdown events} (MT+P\\h \&BC), which introduces Phrase and Bar Countdown events to enhance structural coherence \cite{r26}.
\end{itemize}

\subsubsection{Experiment Configurations.}
We utilize 7280 songs from our STD after data preprocessing, with 90\% selected as training set to train the model and the remaining 10\% as test set to evaluate the performance of the model. The number of layers for both encoder and decoder is 6. The embedding dimension $d_{model}$ is 256, learning rate is 0.001, batch size is 16, and the optimizer we select is Adam with $\epsilon = 10^{-8}$, $\beta_1 = 0.9$, $\beta_2 = 0.999$.

\subsection{Subjective Evaluation}
To assess the quality of the generated music, we conduct a subjective evaluation. Specifically, we invite 10 music experts with professional music training and instrument-playing experience to rate 10 songs generated by STT, three state-of-the-art models and human composers(Ground Truth) on five aspects:

\begin{itemize}[label=\textbullet]
\item \textbf{Melody}: Whether the melody is clear and captivating.
\item \textbf{Rhythm}: Whether the rhythm features consistency.
\item \textbf{Structure}: Whether the melody features a hierarchical structure in its phrases.
\item \textbf{Skeleton}: Whether there are any notes that audibly stand out, playing a role of the musical skeleton.
\item \textbf{Overall}: The overall auditory perception of the entire song.
\end{itemize}

\begin{figure*}[t]
    \centering
    \includegraphics[width=1\textwidth]{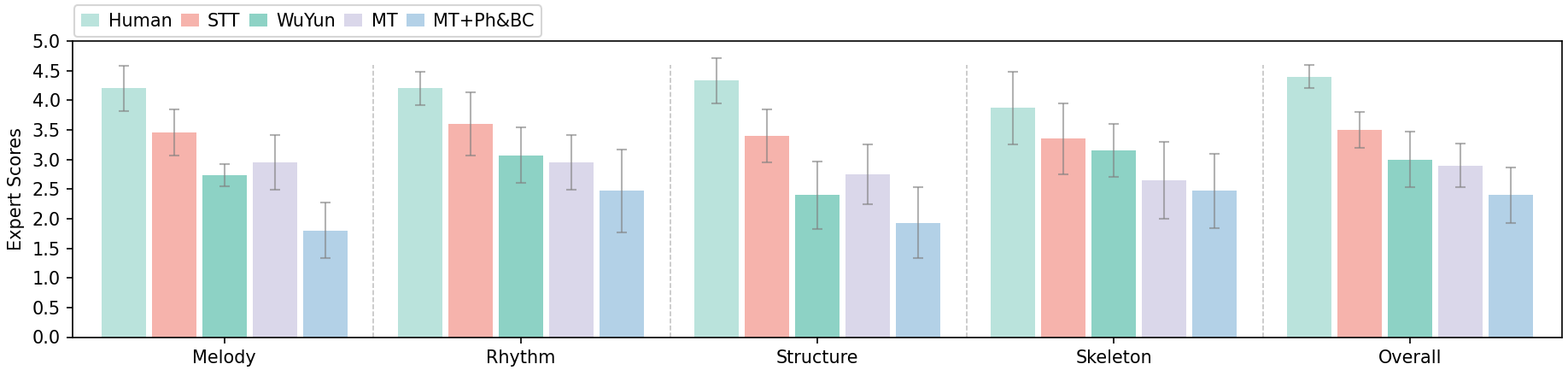}
    \caption{Results of the subjective evaluation. Human, STT, WuYun, MT, MT+Ph\&BC stand for human composition, our proposed model, WuYun architecture, Music Transformer and Music Transformer utilizing Phrase\&Bar Countdown events, respectively.}
    \label{fig:subjective_result}
\end{figure*}

Figure \ref{fig:subjective_result} shows the results of the subjective evaluation. The results indicate that our model, STT, outperforms other state-of-the-art models across all subjective evaluation metrics. This suggests that STT is capable of generating melodies that are more captivating, structures that are clearer, and themes that are more consistently coherent compared to other models. Specially, experts note that STT exhibits a more prominent hierarchical structure in the generated melody compared to other baseline models. However, compared to human compositions, the music generated by STT still exhibits some flaws, indicating room for improvement.

\subsection{Objective Evaluation}
To ensure a comprehensive assessment of the generated music, we also perform an objective evaluation using six metrics. Specifically, we propose three objective evaluation mechanisms as follows:

\noindent {\bfseries Theme Rhythm Correspondence (TRC)}

\begin{equation}
    TRC = \min _{i} D_{\text {hamming }}\left(R_{\text {theme }}, R_{i}\right)
\end{equation}
We propose Theme Rhythm Correspondence to evaluate the rhythm of the generated melody in relation to the theme piece. For this study, the first two bars, as prompt during the generation phase, are selected as the theme sequence. $R_{theme}$ is the binary onset vector of the theme piece ($1$ represents an onset, otherwise $0$), similarly, $R_i$ is the binary onset vector of the $i^{th}$ melody with the same length as the theme piece, and $D_{hamming}(R_{theme}, R_i)$ is the hamming distance to compute the difference of the two melodies $R_{theme}$, $R_i$. The smaller the TRC value, the more rhythmically similar the generated melody is to the theme, reflecting better rhythmic coherence.

\noindent {\bfseries Theme Pitch Correspondence (TPC)}
\begin{equation}
    TPC = \min _{i} D_{\text {hamming }}\left(P_{\text {theme }}, P_{i}\right)
\end{equation}
Similarly, we propose Theme Pitch Correspondence to evaluate the generated melody, with $P_{theme}$ and $P_i$ denote the pitch sequence of theme and the $i^{th}$ piece, respectively.

\noindent {\bfseries Pentatonic Scale Consistency (PSC)}
\begin{equation}
S_i =\left\{
\begin{aligned}
& \hfill\phantom{00} 10, && p_i \in \{C,D,E,G,A\} \\
& \hfill\phantom{00} 6, && p_i \in \{F,F^\#,B,B^b,\} \\
& -10, && \hfill\phantom{p \leq r \leq}\hfill\text {others}
\end{aligned}
\right.
\end{equation}

\begin{equation}
    \text{PSC} = \frac{1}{n} \left( \sum_{i=1}^{n} s_i \right)
\end{equation}
We propose Pentatonic Scale Consistency to evaluate the consistency of generated melody in the pitch scale dimension. Traditional Chinese songs are mostly composed using the Chinese Pentatonic Scale, a distinctive system in Chinese music. This scale consists of five tones: $C$, $D$, $E$, $G$, and $A$, which satisfy the perfect fifth intervals. Additionally, four tones ($F$, $F^\#$, $B$ and $B^b$) can be added to play ornamental roles. We rate each note in the melody: assign 10 points if it belongs to $\{C,D,E,G,A\}$, 6 points if it belongs to $\{F,F^\#,B,B^b\}$, and deduct 10 points if it does not adhere to the rules of the pentatonic scale. Finally, compute the average score across all notes to obtain the PSC. PSC evaluates whether a melody follows the pattern of the Chinese pentatonic scale.

Moreover, we utilize Rhythm Consistency (RC), Pitch Entropy (PE) and Pitch Class Entropy (PCE) from MusPy\footnotemark[2] to evaluate the pitch consistency of melody.

\footnotetext[2]{\url{https://salu133445.github.io/muspy/metrics.html}}

\subsubsection{Comparison Result.}
To evaluate the performance, we compare our model, STT, against three baseline models and human compositions (Ground Truth).
Table \ref{lb:comparitive}. shows the result of comparative experiment. STT outperforms other baseline models in all metrics except RC. The closest TPC and TRC values among all baseline models indicate that our model generates more coherent melodies in both pitch and rhythm dimensions. This suggests that Phrase-level Cross-Attention mechanism effectively learns structural features of Small Tunes songs at both macro level and micro level hierarchies. The PSC, PCE and PE value of our model closely match those of ground truth, indicating its capability to generate Small Tunes songs with more consistent melodies. Although STT slightly lags behind the MT model by 1.3\% in the RC metric, its close proximity to the ground truth indicates that both models perform well in generating melodies with consistent rhythm.

\begin{table*}[tbp]
\centering
\caption{Objective evaluation results of comparative experiments. For all metrics, models with values closer to the ground truth demonstrate better performance.}
\setlength{\tabcolsep}{7pt}
\resizebox{0.95\textwidth}{!}{
\begin{tabular}{ccccccc}
\hline
\textbf{Model}           & \textbf{TPC}  & \textbf{TRC}  & \textbf{RC} & \textbf{PSC}  & \textbf{PCE}  & \textbf{PE}   \\
\hline
STT(ours)       & \textbf{4.61} & \textbf{2.67} & 85.2\%   & \textbf{9.75} & \textbf{2.29} & \textbf{2.66} \\
WuYun           & 6.26 & 3.59 & 85.1\%   & 9.78 & 2.28 & 2.62 \\
MT              & 5.39 & 3.2  & 86.2\%   & 9.82 & 2.25 & 2.58 \\
MT+Ph\&BC       & 7.70 & 4.75 & \textbf{86.5}\%   & 9.90 & 2.23 & 2.53 \\
\hline
\multicolumn{1}{c}{Ground Truth} & 3.78 & 1.91 & 87.3\%   & 9.73 & 2.31 & 2.65 \\
\hline
\end{tabular}
}
\label{lb:comparitive}
\end{table*}

\subsubsection{Ablation Result.}
To explore the underlying features of hierarchical structure in the Chinese Small Tunes, we design 6 ablative groups focusing on two key aspects: phrase segmentation and skeleton notes extraction. In addition to the phrase segmentation utilized in our method, we also employ three phrase segmentation methods:
\begin{itemize}[label=\textbullet]
\item No use of phrase segmentation, treating the music sequence as a single segment (abbreviated as No Segment).
\item Selection of 2 bars as the phrase unit, a rule-based approach to phrase segmentation (abbreviated as 2 Bars).
\item Expansion of the phrase boundaries from our phrase segmentation result, combining two phrases into a larger unit (abbreviated as Expansion).
\end{itemize}
Based on these phrase segmentation methods, we additionally design a skeleton notes extraction method, which reduces the number of extracted skeleton notes by randomly removing 50\% skeleton notes within each phrase.

\begin{table*}[tbp]
\centering
\caption{Objective evaluation results of ablation experiments. \textit{Phrase} and \textit{Skeleton} are abbreviations for the ablation methods of phrase segmentation and skeleton notes extraction, respectively. For all metrics, models with values closer to the ground truth demonstrate better performance.}
\setlength{\tabcolsep}{3pt}
\resizebox{1\textwidth}{!}{
\begin{tabular}{ccccccccc}
\hline
\textbf{Group}   & \textbf{Phrase}  & \textbf{Skeleton} & \textbf{TPC}           & \textbf{TRC}           & \textbf{RC}        & \textbf{PSC}           & \textbf{PCE}           & \textbf{PE}            \\
\hline
1(STT)   & - & -  & \textbf{4.61} & \textbf{2.67} & 85.2\%          & \textbf{9.75} & 2.29          & \textbf{2.66} \\
2(WuYun) & No Segment       & -  & 6.26          & 3.59          & 85.1\%          & 9.78          & 2.28          & 2.62          \\
3        & 2 Bars           & -  & 6.58          & 3.76          & 84.5\%          & 9.76          & \textbf{2.30} & 2.67          \\
4        & Expansion        & -  & 5.11          & 3.17          & 85.0\%          & 9.80          & 2.26          & 2.60          \\
5        & - & Remove 50\%       & 4.94          & 2.92          & 85.5\%         & 9.80          & 2.25          & 2.57          \\
6        & No Segment       & Remove 50\%       & 5.25          & 3.18          & 85.1\%          & 9.81          & 2.24          & 2.55          \\
7        & 2 Bars           & Remove 50\%       & 7.21          & 4.02          & 84.9\%          & 9.77          & 2.29          & 2.62          \\
8        & Expansion        & Remove 50\%       & 5.23          & 3.12          & \textbf{85.8}\% & 9.82          & 2.24          & 2.56          \\
\hline
\multicolumn{3}{c}{Ground Truth}                & 3.78          & 1.91          & 87.3\%          & 9.73          & 2.31          & 2.65         \\
\hline
\end{tabular}
}
\label{lb:abaltion}
\end{table*}

Table \ref{lb:abaltion} shows the result of ablation experiment. We construct 6 ablation groups (Group 2-8) by adjusting the phrase segmentation and skeleton notes extraction strategies. Group 1-4 and 5-8 each employ the same skeleton notes extraction strategy within their groups but utilize different phrase segmentation strategies. Group 1 and Group 5 respectively achieve the best performance in TPC and TRC, indicating that our phrase segmentation strategy contributes to generating coherent melodies. Furthermore, Group 1 outperforms Group 2-4 in almost all metrics except PCE, suggesting that inappropriate segment boundaries are detrimental to capturing the structural features within Small Tunes songs. Moreover, Group 1 outperforms Group 5, indicating that an appropriate number of skeleton notes contribute to guiding the melody generation and constructing the hierarchical structure.

\section{CONCLUSION}
In order to study the hierarchical structural features within music, we delve into multi-level hierarchies: at the macro-level hierarchy, we conduct phrase segmentation algorithm to study the impact of phrase on the overall structural organization, and at the micro-level hierarchy, we design a skeleton notes extraction strategy to explore how skeleton notes within phrases influence the melody generation. Building upon this, we propose a novel Phrase-level Cross-Attention to capture the intrinsic relationship among multi-level hierarchies. Moreover, we train our proposed model: Small Tunes Transformer on our own established dataset: Small Tunes Dataset, providing a new perspective for the composition of Chinese-style music. We design three novel metrics to evaluate music from rhythm and melody dimensions. The experiment results indicate that our model outperforms other state-of-the-art models on both subjective and objective evaluations. Additionally, we add several ablative groups to deeply explore the intrinsic features within hierarchical structures. In future work, we aim to extend our study of macro and micro-level hierarchies within music, particularly focusing on polyphonic compositions.

\bibliographystyle{splncs04}
\bibliography{reference}

\end{document}